\documentclass[aps,prb,showpacs,tightenlines,twocolumn,nofootinbib,nobibnotes,superscriptaddress]{revtex4-1}
\usepackage{amsmath,amssymb,amsfonts,bm}
\usepackage{graphicx}
\usepackage{epstopdf}
\usepackage{dcolumn}
\usepackage{mathrsfs}
\usepackage{bbold}
\usepackage{dsfont}
\usepackage{ulem}
\usepackage[colorlinks=true,linkcolor=blue,citecolor=blue, urlcolor=blue,bookmarks=false]{hyperref}
\begin{document}
\title{Phase diagrams of Weyl semimetals with competing intraorbital and interorbital disorders}
\date{\today }
\author{Rui Chen}
\affiliation{Department of Physics, Hubei University, Wuhan 430062, China}
\author{Chui-Zhen Chen}
\affiliation{Department of Physics, Hong Kong University of Science and Technology, Clear Water Bay, Hong Kong, China}
\author{Jin-Hua Sun}
\affiliation{Department of Physics, Ningbo University, Ningbo 315211, China}
\author{Bin Zhou}\thanks{binzhou@hubu.edu.cn}
\author{Dong-Hui Xu} \thanks{donghuixu@hubu.edu.cn}
\affiliation{Department of Physics, Hubei University, Wuhan 430062, China}

\begin{abstract}
A Weyl semimetal (WSM) is a topological material that hosts Weyl fermions as quasiparticles in the bulk. We study the combined effect of intra- and interorbital disorders on WSMs by adopting a tight-binding model that supports the WSM, three-dimensional quantum anomalous Hall insulator (3D QAHI) and normal insulator (NI) phases in the clean limit. Based on the calculation of the localization length and the Hall conductivity, we present rich phase diagrams due to the interplay of intra- and interorbital disorders. We find that the WSM with well-separated Weyl nodes is stable to both weak intra- and interorbital disorders. However, weak intraorbital disorder can gap out a WSM close to the 3D QAHI phase in the clean phase diagram, forming a 3D QAHI, and it can also drive a NI near the WSM phase to a WSM. By contrast, weak interorbital disorder can cause a 3D QAHI-WSM transition for a 3D QAHI in proximity to the WSM phase in the clean limit, and it can annihilate a WSM near the NI phase, bringing about a WSM-NI transition. We observe a diffusive anomalous Hall metal (DAHM) phase at moderate disorder strength. The DAHM appears in a wide range of the phase diagram when the intraorbital disorder dominates over the interorbital disorder, while the DAHM is found to exist in a narrow region or be missing in the phase diagram when the interorbital disorder dominates.
\end{abstract}

\maketitle

\section{Introduction}
Since the discovery of topological insulators \cite{Hasan10RMP,Qi11RMP}, the search for exotic topological phases of matter has been a major research topic in condensed matter physics. Weyl semimetals \cite{Wan11PRB} (WSMs) are a newly discovered class of gapless topological phases, which are characterized by linearly dispersive band-touching points, called Weyl nodes. The low energy excitations in the vicinity of a Weyl node behave exactly as Weyl fermions. Each Weyl node can be viewed as a monopole or an anti-monopole of the Berry curvature in momentum space. The Weyl nodes in a WSM must come in pairs with opposite chirality according to the no-go theorem \cite{Nielsen81PhysicsB1,Nielsen81PhysicsB2}, and are protected by topology. Owing to the nontrivial band topology, WSMs exhibit Fermi arc electron states connecting Weyl nodes on the surface. Because of the existence of Weyl nodes and Fermi arc states, WSMs exhibit exotic transport properties such as the chiral anomaly induced negative magnetoresistance\cite{XiongJ15Science,HuangX15PRX,ZhangCL16NatComm} and the Weyl orbit physics \cite{Potter14NatComm,Moll16Nature,DaiX16NatPhys,ZhangCL17NatPhys}.

In the last few years, the effect of the chemical potential disorder in WSMs has been intensively studied\cite{Roy16arXiv,Hosur12PRL,Ominato14PRB,Kobayashi2014,Biswas14PRB,Sbierski14PRL,Sbierski15PRB,Baireuther14PRB,Nandkishore14PRB,Pixley15PRL,Pixley16PRB,Pixley16PRX,Altland15PRL,Syzronov15PRL,Syzronov18ARCM,Chen15PRL,Liu16PRL,Shapourian16PRB,Bera16PRB}. It has been found that the WSM phase is robust against weak chemical potential disorder. Remarkably, a normal insulator (NI) can be converted into a WSM at finite disorder strength. The chemical potential disorder can also open a band gap in WSMs, resulting in a WSM-three dimensional quantum anomalous Hall insulator (3D QAHI) phase transition \cite{Chen15PRL,Liu16PRL,Shapourian16PRB}. By means of the self-consistent Born approximation, the phase transitions at weak disorder can be understood by the renormalization effect of the topological mass. The same effect has also been found in type-II WSMs with weak disorders very recently\cite{Park17PRB,Wu17PRB}. Interestingly, it was found that the WSM becomes a diffusive metal with finite Hall conductance at moderate disorder \cite{Pixley15PRL,Kobayashi2014,Sbierski14PRL,Sbierski15PRB,Pixley16PRB,Pixley16PRX,Chen15PRL,Shapourian16PRB,Liu16PRL,Bera16PRB}, which is called the diffusive anomalous Hall metal (DAHM)\cite{Chen15PRL}. 

Except for the chemical potential disorder, there also exists the interorbital disorder, which has been proposed to play a distinct role in modulating topological properties of various systems \cite{Song12PRB,Hu16PRB,QiaoZ16PRL,Hung16PRB}. For example, the chemical potential disorder can induce a phase transition from a NI to the topological Anderson insulator \cite{LiJ09PRL,Jiang09PRB,Groth09PRL,Jiang16cpb}, whereas for the interorbital disorders, the topological Anderson insulator phase is missing \cite{Song12PRB}. Although the effect of the chemical potential disorder on WSMs is well established, the stability of WSMs in the presence of the interorbital disorder has not been well explored.

 In this paper, we investigate the effect of coexisting intra- and interorbital disorders in WSMs by employing both numerical and analytical methods. The competition of intra- and interorbital disorders gives rise to rich phase diagrams which are determined by numerically computing the normalized localization length and the Hall conductivity. We show that a WSM is robust against weak disorders when the Weyl nodes are well separated in momentum space. Weak intraorbital disorder may annihilate a WSM close to the 3D QAHI phase in the clean phase diagram, causing a 3D QAHI, while it could drive a NI near the WSM phase to a WSM. By contrast, weak interorbital disorder can destroy a 3D QAHI in proximity to the WSM phase in the clean limit, resulting in a 3D QAHI-WSM transition, or it can gap out a WSM close to the NI phase, leading to a WSM-NI transition. These phase transitions at weak disorder are analytically verified by using the self-consistent Born approximation and can be explained by the topological mass renormalization. Before Anderson localization induced by strong disorder occurs, we observe a 3D DAHM phase at moderate disorder strength. The DAHM appears in a wide range of the phase diagram when the intraorbital disorder dominates over the interorbital disorder, while the DAHM is found to exist in a narrow region or be missing in the phase diagram when the interorbital disorder dominates.

 The rest of the paper is organized as follows. In Sec.~\ref{Model}, we introduce a disordered two-band WSM Hamiltonian and then give the details of our numerical methods. Then we provide phase diagrams in $(W_{xy},W_z)$ space for the disordered 3D QAHI, WSM and NI phases, respectively, in Sec.~\ref{Numerical}, and interpret the weak disorder results in terms of the self-consistent Born approximation in Sec.~\ref{SecBorn}. At last, a brief discussion and summary is presented in Sec.~\ref{Conclusion}.
\section{Model and method}

We begin with a simple two-band tight-binding model Hamiltonian defined on a cubic lattice with unity lattice constant
 \cite{Yang11PRB}
\begin{align}
H^{0}\left( \mathbf{k}\right)   =&\left(  m_{z}-t_{z}\cos k_{z}\right)
\sigma_{z}+m_{0}\left(  2-\cos k_{x}-\cos k_{y}\right)  \sigma_{z}\nonumber\\
&  +t_{x}\sigma_{x}\sin k_{x}+t_{y}\sigma_{y}\sin k_{y}, \label{Hk}%
\end{align}
where $t_{i}(i=x,y,z)$, $m_{z}$, and $m_{0}$ are the model parameters. $\sigma_{i}$ are Pauli matrices acting on the orbital (or spin) space. This Hamiltonian lacks the time-reversal symmetry, and can be considered as a stack of 2D QAHIs with an orbital dependent tunneling $t_z$.

\label{Model}
\begin{figure}[hptb]
	\includegraphics[width=8cm]{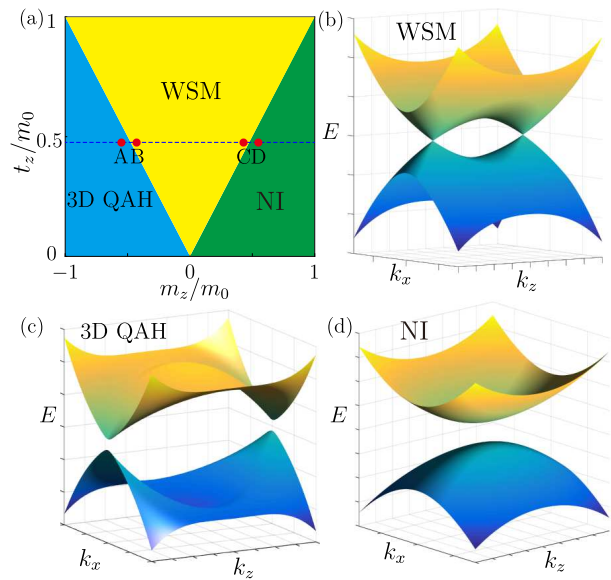} \caption{(Color online)(a) Phase diagram of the pure Weyl semimetal Hamiltonian $H_{0}\left(  \mathbf{k}\right) $ on the
		$t_{z}/m_{0}$-$m_{z}/m_{0}$ plane. The regions colored with blue, yellow, and green correspond to the 3D QAHI, WSM, and NI phases. The red points correspond to $m_{z}/m_{0}=-0.55, -0.45, 0.45$, and $0.55$. (b)-(d) Bulk spectra of the WSM, 3D QAHI, and NI phases at $k_y=0$.}%
	\label{fig1}%
\end{figure}
The phase diagram of the pure Hamiltonian $H^{0}\left(  \mathbf{k}\right)
$ is shown in Fig.~\ref{fig1}(a). When $| m_{z}/m_{0}|< t_{z}/m_{0}$, the model yields a pair of gapless Weyl nodes at $\left( 0, 0, \pm k_z^0 \right) $ with $k^0_z=\arccos{(m_{z}/t_{z})}$, which realizes the WSM phase [Fig.~\ref{fig1}(b)]. For $-2+t_z/m_0<m_{z}/m_{0}<-t_{z}/m_{0}$, the model has a topological bulk gap, corresponding to the 3D QAHI phase [Fig. \ref{fig1}(c)]. When $m_{z}/m_{0}>t_{z}/m_{0}$, the bulk gap of the model is topologically trivial, giving rise to the NI phase [Fig. \ref{fig1}(d)]. In the following calculations, we fix the parameters as $t_{x}=t_{y}=1$, $m_{0}=2.1$ and the chemical potential $\mu=0$.

Now we introduce the intra- and interorbital disorders to the model, and the onsite disorder terms write
\begin{equation}
\Delta H=\sum_{\mathbf{i}}\left[  U_{x}^{\mathbf{i}}
c_{\mathbf{i}}^{\dag}\sigma_{x}c_{\mathbf{i}}+U_{y}^{\mathbf{i}}
c_{\mathbf{i}}^{\dag}\sigma_{y}c_{\mathbf{i}}+U_{z}^{\mathbf{i}}
c_{\mathbf{i}}^{\dag}\sigma_{z}c_{\mathbf{i}}  \right],
\end{equation}
where $c^\dagger_{\mathbf{i}} (c_{\mathbf{i}})$ represents the creation (annihilation) electron operator at site $\mathbf{i}$, the first two terms describe the interorbital disorder, and the last term represents the intraorbital disorder. $U_{x,y,z}^{\mathbf{i}}$ are uniformly distributed within $\left[  -W_{x,y,z}/2,W_{x,y,z}/2\right]$ with $W_{x,y,z}$ representing the disorder strengths.
We take $W_x=W_y=W_{xy}$ in our numerical simulation since both $U_{x}^{\mathbf{i}}$ and $U_{y}^{\mathbf{i}}$ describe the interorbital disorder and play the same role in phase transition and localization. It is necessary to point out that the chemical potential disorder (the $\sigma_0$-type intraorbital disorder), which has been intensively studied in the literature\cite{Shapourian16PRB,Chen15PRL,Liu16PRL}, is ignored in our model. The reason is that the $\sigma_0$-type and $\sigma_z$-type disorders give the same self-energy correction to the topological mass within the self-consistent Born approximation theory, and they are equivalent in determining phase transitions. Thus, the total disordered Hamiltonian is $H=H^{0}+\Delta H$.

We will construct phase diagrams where each phase is determined by the localization length $\lambda$ and the Hall conductivity $\sigma_{xy}$. To calculate the localization length, we use the standard transfer matrix method\cite{MacKinnon81PRL,MacKinnon83PhysB,Kramer93RPP,Yamakage13PRB,Chen15PRB,Nagaosa03PRL,Slevin04PRB,Onoda07PRL,Kobayashi13PRL}, and consider a quasi-one-dimensional system of cross section $L_{x} \times L_{y}=L \times L$ and length $L_{z}$ with periodic boundary conditions in the $x$ and $y$ directions. The system can be divided into slices along the $z$ direction, and the Schr\"odinger equation is expressed as
\begin{equation}
H_{n,n}\psi_{n}+H_{n,n+1}\psi_{n+1}+H_{n,n-1}\psi_{n-1}=E\psi_{n},
\end{equation}
where $\psi_{n}$ is the wave function at the $n$th slice, $H_{n,n}=\left<n\right|H\left|n\right>$, $H_{n,n+1}=\left<n\right|H\left|n+1\right>$, and $H_{n,n-1}=\left<n\right|H\left|n-1\right>$. The Schr\"odinger equation can be rearranged into the following form in terms of the transfer matrix $T_n$: \begin{equation}
\begin{pmatrix}
\psi_{n+1}\\
\psi_{n}%
\end{pmatrix}
=T_{n}%
\begin{pmatrix}
\psi_{n}\\
\psi_{n-1}%
\end{pmatrix},
\end{equation}
with
\[
T_{n}=%
\begin{pmatrix}
H_{n,n+1}^{-1}(E\mathds{1}-H_{n,n}) & -H_{n,n+1}^{-1}H_{n,n-1}\\
\mathds{1} & 0
\end{pmatrix}.
\]
A Lyapunov exponent $\gamma^i$ is defined by the following limiting value
\begin{equation}
\gamma^i=\lim_{L_z\rightarrow\infty}\frac{\ln \varepsilon_i}{2L_z},
\end{equation}
where $\varepsilon_i$ is the $i$-th positive eigenvalue of the matrix product $T^\dagger T$, and $T=\prod\nolimits_{n=1}^{L_z}T_{n}$. The smallest positive Lyapunov exponent $\gamma_{\text{min}}$ is related to the localization length $\lambda$ by
$\lambda=1/\gamma_{\text{min}}$.
Furthermore, the normalized localization is defined as $\Lambda\equiv\lambda/L$. In general, the normalized localization length $\Lambda$ increases with $L$ in a metallic phase, and decreases with $L$ in an insulator phase. At the critical point of phase transition, $\Lambda$ is independent of $L$.

In the presence of intra- and interorbital disorders, we calculate the ensemble averaged Hall conductivity $\sigma_{xy}$ of 3D samples by use of the noncommutative Kubo formula under periodic boundary conditions \cite{Prodan11JPA,Prodan13AMRE,Song14PRB1,Song14PRB2}. In the clean limit, for fixed $k_z$, one can treat the Hamiltonian as a 2D Dirac Hamiltonian $H^{0}_{k_z}\left(k_x,k_y\right)$ with a $k_z$-dependent mass term \cite{Burkov11PRL,Burkov11PRB}. In the present model, the Weyl nodes are located at $\left(0,0,\pm k_z^0\right)$. For $k_z<|k_z^0|$, each of the topological nontrivial 2D Hamiltonian $H^{0}_{k_z}\left(k_x,k_y\right)$ contributes a quantized Hall conductivity $\sigma_{xy}^{2D}\left( k_z\right)=e^2/h$. Therefore, the total 3D Hall conductivity of $H^{0}\left( \mathbf{k}\right)$ is given by $\sigma_{xy}=\sum_{k_z} \sigma_{xy}^{2D}(k_z)/L_z$, which is proportional to the distance between the two Weyl nodes.

\section{Numerical Simulation}
\label{Numerical}
In this section, we investigate the combined effect of intra- and interorbital disorders in the localization effect and phase transitions. We map out phase diagrams in the parameter space ($W_{xy}$, $W_z$) for the disordered 3D QAHI, WSM, and NI phases in Secs.~\ref{3DQAHI}, \ref{WSM}, and \ref{NI}, respectively.
\subsection{Disordered 3D QAHI}
\begin{figure}[hptb]
	\includegraphics[width=8cm]{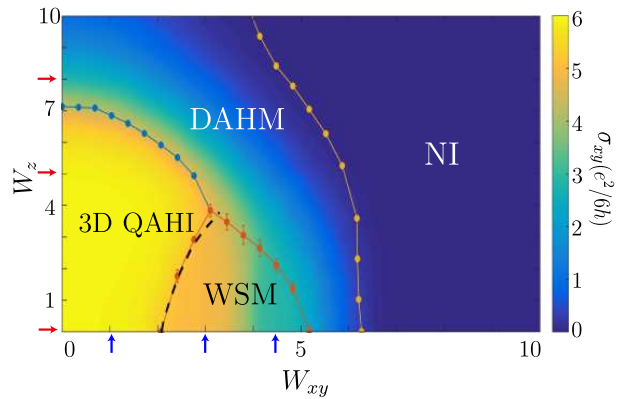} \caption{(Color online) Phase diagram in the ($W_{xy}$, $W_{z}$) space for the disordered 3D QAHI with $m_{z}/m_{0}=-0.55$. The color map represents the Hall conductivity
		$\sigma_{xy}$ and the solid lines are determined by the localization length. The system size for $\sigma_{xy}$ is $20 \times 20 \times 6$. The black dashed line is obtained from the self-consistent Born approximation method, which corresponds to the phase boundaries defined as $\Delta_{\tilde{E}}\left(0,0,\pi\right)=0$. }%
	\label{fig2}%
\end{figure}

First of all, we study the disorder effect of a 3D QAHI near the phase boundary between 3D QAHI and WSM in the clean phase diagram [marked by the red dot A in Fig.~\ref{fig1}(a)]. The phase diagram for the disordered 3D QAHI is plotted in the space of inter- and intraorbital disorder strengths ($W_{xy}$, $W_z$) in Fig.~\ref{fig2}. The color map shows the Hall conductivity $\sigma_{xy}$ that distinguishes the 3D QAHI, the WSM and the 3D DAHM phases. The solid lines indicate the phase boundaries determined by the scale-invariant points of the renormalized localization length $d\Lambda/dL=0$.

\label{3DQAHI}
\begin{figure}[hptb]
	\includegraphics[width=8cm]{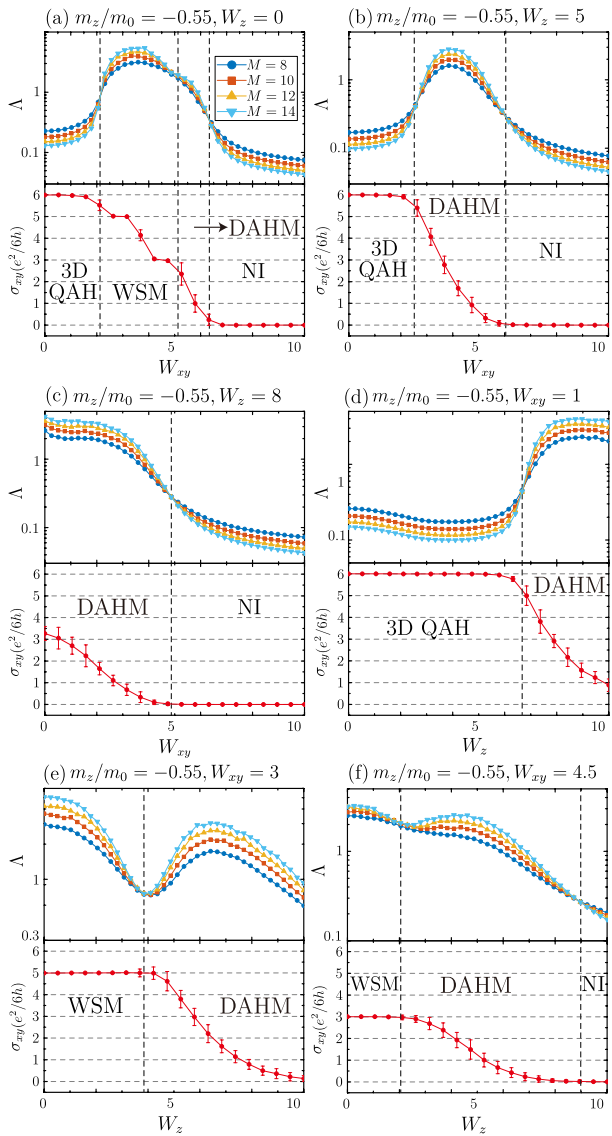} \caption{(Color online) The normalized localization length and Hall conductivity as a function of $W_{xy}$ for (a) $W_{z}=0$, (b) $W_{z}=5$, (c) $W_{z}=8$, and of $W_{z}$ for (d) $W_{xy}=1$, (e) $W_{xy}=3$, (f) $W_{xy}=4.5$, which correspond to the red and blue arrows shown in Fig.~\ref{fig2}, respectively. We set the parameter $m_z/m_0=-0.55$. The cross-section sizes in calculating the normalized localization length are $L=8$ (blue circle), 10 (red square), 12 (orange up triangle), and 14 (cyan down triangle). The dashed vertical lines represent the phase transition lines. The system size for $\sigma_{xy}$ is $30 \times 30 \times 6$. The error bars are magnified three times to show the standard deviation of the conductivity for 100 samples.}%
	\label{fig3}%
\end{figure}
Interestingly, we find a WSM phase emerges between the 3D QAHI and DAHM phases only when the interorbital disorder $W_{xy}$ dominates over the  intraorbital disorder $W_z$. On the other hand, the disordered 3D QAHI always goes into the DAHM phase before it is finally localized by the intra- and interorbital disorders. Note that we do not show the localized NI region in the phase diagram for weak $W_{xy}$, as it requires pretty strong intraorbital disorders (for example, the critical disorder strength $W^c_z\approx 20 $ at $W_{xy}=0$).

To gain further insight into the competition between $W_{xy}$ and $W_z$, we take a few line cuts of Fig.~\ref{fig2} indicated by the arrows, and show them in Fig.~\ref{fig3}.
Figures~\ref{fig3}(a)-\ref{fig3}(c) display $\Lambda$ and $\sigma_{xy}$ of disordered 3D QAHI as a function of $W_{xy}$ for different $W_z$, respectively. When $W_z=0$, as shown in Fig.~\ref{fig3}(a), the 3D QAHI remains stable at weak $W_{xy}$, which is characterized by $d\Lambda/dL<0$ and a fully quantized Hall conductivity $\sigma_{xy}=e^2/h$. With increasing $W_{xy}$, a phase transition occurs at the critical point where $d\Lambda/dL=0$, beyond which a metal phase emerges as $d\Lambda/dL>0$. At the same time, $\sigma_{xy}$ decreases from a fully quantized conductivity $e^2/h$ to a fractional conductivity $\sigma_{xy}=5e^2/6h$. Thus, we identify this metal phase as a WSM phase. Further increasing $W_{xy}$, the Hall conductivity reduces to $\sigma_{xy}=3e^2/6h$. It is worth noting that the appearance of discrete fractional $\sigma_{xy}$ is due to the finite size effect.
The decrease of the Hall conductivity in the WSM phase can be understood within the framework based on the self-consistent Born approximation (see Sec. \ref{SecBorn}). In this effective physical picture, increasing the interorbital disorder strength reduces the distance between the Weyl nodes and decreases the Hall conductivity as it is proportional to the distance. This is in striking contrast to the case of the chemical potential disorder, in which the Hall conductivity rises with increasing the disorder strength \cite{Shapourian16PRB,Chen15PRL,Liu16PRL}.
Finally, before the system is localized by strong disorder, $\sigma_{xy}$ starts to exhibit fluctuations, indicating an intermediate DAHM phase appears between the WSM and NI phases. However, for finite $W_z$, the results in Figs. \ref{fig3}(b) and \ref{fig3}(c) show different features from those in Fig.~\ref{fig3}(a) with $W_z=0$. At $W_{z}=5$ [Fig.~\ref{fig3}(b)], the 3D QAHI phase is still stable, and no WSM phase appears as $W_{xy}$ increases. That is because the intraorbital disorder drives the 3D QAHI away from the phase boundary between 3D QAHI and WSM. For a larger strength $W_{z}=8$ [Fig. \ref{fig3}(c)], the system enters into the DAHM phase even at $W_{xy}=0$, meanwhile, the anomalous Hall conductivity maintains finite until $W_{xy}=5$.

We also plot the localization length and the Hall conductivity as a function of $W_{z}$ at different $W_{xy}$ in Figs.~\ref{fig3}(d)-\ref{fig3}(f). For different $W_{xy}$, the system undergoes distinct phase transitions with increasing $W_{z}$, including a 3D QAHI-DAHM transition in Fig.~\ref{fig3}(d), and phase transitions from the WSM phase to the DAHM phase in Figs.~\ref{fig3}(e) and \ref{fig3}(f). The phase diagram in Fig.~\ref{fig2} is obtained by repeating these procedures at different values of $W_{xy}$ and $W_{z}$.
\begin{figure}[hptb]
\includegraphics[width=8cm]{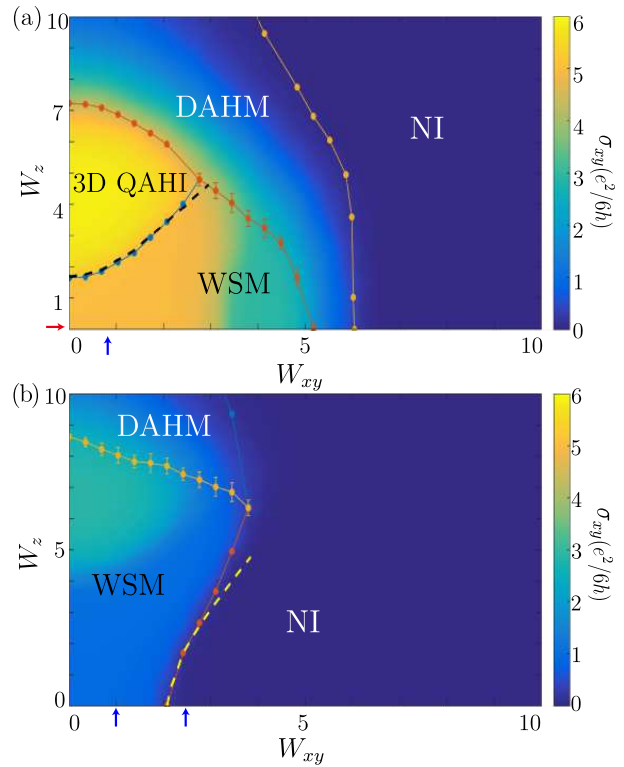} \caption{(Color online) Phase diagrams in the ($W_{xy}$, $W_{z}$) space for the disordered WSM with (a) $m_{z}/m_{0}=-0.45$ and (b) $m_{z}/m_{0}=0.45$. The color map represents the Hall conductivity
$\sigma_{xy}$ and the solid lines are determined by the localization length $\Lambda$. The system size for $\sigma_{xy}$ is $20 \times 20 \times 6$. The black and yellow dashed lines in (a) and (b) are obtained from the self-consistent Born approximation, which correspond to the phase boundary defined as $\Delta_{\tilde{E}}\left(0,0,\pi\right)=0$ and $\Delta_{\tilde{E}}\left(0,0,0\right)=0$, respectively.}%
\label{fig4}%
\end{figure}
\subsection{Disordered WSM}
\label{WSM}

\begin{figure}[hptb]
\includegraphics[width=8cm]{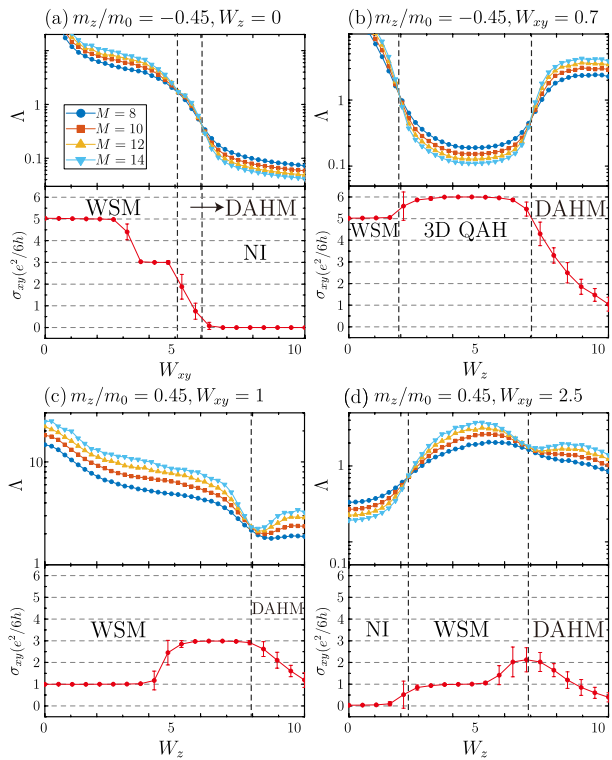} \caption{(Color online) The normalized localization length and Hall conductivity as a function of $W_{xy}$ for (a) $m_z/m_0=-0.45,W_{z}=0$, and of $W_{z}$ for (b) $m_z/m_0=-0.45,W_{xy}=0.7$, (c) $m_z/m_0=0.45,W_{xy}=1$, (d) $m_z/m_0=0.45,W_{xy}=2.5$, which correspond to the red and blue arrows shown in Fig.~\ref{fig4}, respectively. The cross-section sizes in calculating the normalized localization length are $L=8$ (blue circle), 10 (red square), 12 (orange up triangle), and 14 (cyan down triangle). The dashed vertical lines represent the phase transition lines. The system size for $\sigma_{xy}$ is $30 \times 30 \times 6$. The error bars are magnified three times to show the standard deviation of the conductivity for 100 samples. }%
\label{fig5}%
\end{figure}

In this section, we study the competition of intra- and interorbital disorders in WSMs. We consider two WSMs with $m_z/m_0=-0.45$ [see the red dot B in Fig.~\ref{fig1}(a)] and $m_z/m_0=0.45$ [see the red dot C in Fig.~\ref{fig1}(a)], which are near the phase boundaries of 3D QAHI-WSM and WSM-NI, respectively. Figures~\ref{fig4}(a)-\ref{fig4}(b) show two corresponding phase diagrams of disordered WSMs in the ($W_{xy}$, $W_z$) space. In Fig.~\ref{fig4}(a), we find that a 3D QAHI phase appears between the WSM and the DAHM phases only when the intraorbital disorder strength $W_{z}$ dominates over the interorbital disorder strength $W_{xy}$.
Like the disordered 3D QAHI case, the DAHM phase emerges before the system becomes localized at strong disorders.
On the contrary, in Fig.~\ref{fig4}(b), the 3D QAHI phase is absent, and only the DAHM phase appears when $W_z$ is much greater than $W_{xy}$. The system undergoes a direct phase transition from a WSM to a NI when the interorbital disorder dominates. Taking a few line cuts of Fig.~\ref{fig4} marked by the arrows, we plot $\Lambda$ and $\sigma_{xy}$ of as a function of disorder strength $W_{xy}$ or $W_z$ in Fig.~\ref{fig5}. At $W_z=0$, as shown in Fig.~\ref{fig5}(a), the interorbital disorder induces a phase transition from the WSM phase to the DAHM phase at $W_{xy}\approx 5$. Before that, the Hall conductivity decreases from $5e^2/6h$ to $3e^2/6h$ with increasing the interorbital disorder. This is consistent with the discussion of the interorbital disorder in the previous section. On the other hand, in Fig.~\ref{fig5}(b), when we fix $W_{xy}=0.7$ and increase $W_z$, a phase transition from the WSM to the 3D QAHI occurs as evidenced by a Hall conductivity transition from $5e^2/6h$ to $e^2/h$. This means that when the intraorbital disorder dominates, the WSM phase tends to move towards the phase boundary of 3D QAHI-WSM, which is similar to the case of the $\sigma_0$-type disorder\cite{Shapourian16PRB,Chen15PRL,Liu16PRL}. Figure~\ref{fig5}(c) shows that, for a fixed $W_{xy}=1$, a phase transition from the WSM to the DAHM happens at $W_z=8$. Before the phase transition occurs, we can see that $\sigma_{xy}$ increases at $W_z\approx5$ and reaches $3e^2/6h$ before decreasing at $W_z=8$. Figure~\ref{fig5}(d) shows the multiple NI-WSM-DAHM phase transition as a function of $W_z$ for a fixed $W_{xy}=2.5$. Surprisingly, for $W_{xy}=2.5$, the Hall conductivity even disappears at vanishing $W_z$ since the WSM phase collapses, however it reappears as $W_z$ increases. The reentrant behavior of the WSM phase originates from the interplay of intra- and interorbital disorders.

\subsection{Disordered NI}
\label{NI}

\begin{figure}[hptb]
	\includegraphics[width=8cm]{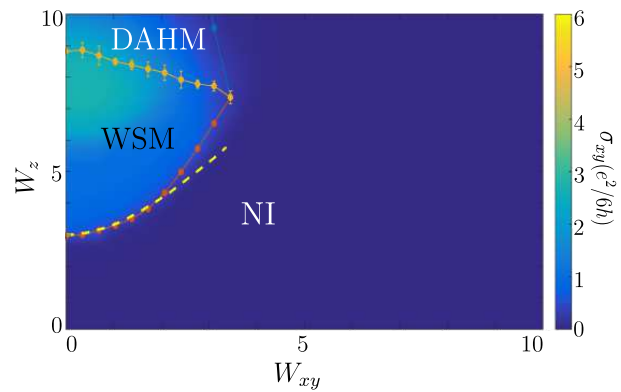} \caption{(Color online) Phase diagram in the ($W_{xy}$, $W_{z}$) space for the disordered NI with $m_{z}/m_{0}=0.55$. The color map represents the Hall conductivity
		$\sigma_{xy}$ and the solid lines are determined by the localization length $\Lambda$. The system size for $\sigma_{xy}$ is $20 \times 20 \times 6$. The yellow dashed line is obtained from the self-consistent Born approximation, which corresponds to the phase boundary defined as $\Delta_{\tilde{E}}\left(0,0,0\right)=0$.}%
	\label{fig6}%
\end{figure}

The phase diagram of the disordered NI is shown in Fig. \ref{fig6}. The location of NI in the clean phase diagram Fig.~\ref{fig1}(a) is marked by red dot D. We find that a WSM phase emerges in the phase diagram when the intraorbital disorder is dominant, while the NI phase occupies the phase diagram when the interorbital disorder dominates.
\section{Born approximation}
\label{SecBorn}
To better understand the phase transitions induced by the intra- and interorbital disorders at weak disorder strength, we deal with them within the self-consistent Born approximation in which high-order scattering processes are neglected. In the basis of the pure Hamiltonian $H^0(\mathbf{k})$, the self-energy correction $\Sigma$ induced by intra- and interorbital disorders are given by the following integral equation\cite{Groth09PRL,Song12PRB,Ominato14PRB,Hung16PRB}:
\begin{equation}
\Sigma(E)=\sum_{i=x,y,z}\frac{W_{i}^{2}}{12}\left(  \frac{a}{2\pi}\right)  ^{3}\int_{\text{FBZ}}%
d\mathbf{k}[\sigma_i(E-H^0\left(  \mathbf{k}\right)  -\Sigma
)^{-1}\sigma_i]\text{,}%
\label{Born}
\end{equation}
where the self-energy can be expressed as $\Sigma=\Sigma_0\sigma_0+\Sigma_x\sigma_x+\Sigma_y\sigma_y+\Sigma_z\sigma_z$. The coefficient $1/12$ comes from the variance $\left\langle U_i^{2}\right\rangle =W_i^{2}/12$ of a random variable uniformly distributed in the range $\left[-W_i/2,W_i/2\right]$.
This integration is over the first Brillouin zone (FBZ). By calculating the band gap $\Delta_{\tilde{E}}\left( \mathbf{k}\right)$ of the renormalized Hamiltonian $\tilde{H}^0\left(  \mathbf{k}\right)$=$H^0\left( \mathbf{k}\right)+\Sigma$ as functions of $W_{xy}$ and $W_{z}$, we obtain the curves in Figs.~\ref{fig2}, \ref{fig4}, and \ref{fig6}.
$\Delta_{\tilde{E}}\left( \mathbf{k}\right)$ is defined as $\Delta_{\tilde{E}}\left( \mathbf{k}\right)=E_{c}(\mathbf{k})-E_{v}(\mathbf{k})$ where $E_{c}$ and $E_{v}$ are the energy spectra of the conduction and valence bands. The black dashed lines in Figs. \ref{fig2} and \ref{fig4}(a) correspond to the phase transition lines $\Delta_{\tilde{E}}\left(0,0,\pi\right)=0$, which means the gap closes at the Brillouin zone boundary. The yellow dashed lines in Figs. \ref{fig4}(b) and \ref{fig6} correspond to gap closure at the Brillouin zone center with $\Delta_{\tilde{E}}\left(0,0,0\right)=0$. We can see that the results based on the Born approximation fit well with the numerical ones. Both intra- and interorbital disorders have a renormalization effect on the bulk band gap $\Delta_{\tilde{E}}$, leading to various topological phases and exotic phase transitions in the present system.

To corroborate the physical interpretation of the numerical simulation, we neglect $\Sigma$ on the right hand side of Eq.~(\ref{Born}) and expand $H^0$ to the $k^2$ order. We find that the intra- and interorbital disorders renormalize the model parameter $m_z$ with
\begin{equation}
\tilde{m}_{z}=m_{z}+\left(2W_{xy}^{2}-W_{z}^{2}\right)T,
\end{equation}
where
\begin{equation}
T= \frac{a^{2}}{96\pi ^{2}m_{0}}\int_{-\pi}^{\pi} \ln  \left|\left( \frac{m_0}{t_z-t_z k_z^2-m_z}\right)^2 \left( \frac{\pi}{a} \right)^4\right |dk_z,
\end{equation}
and we keep only the logarithmically divergent part of the integral. Clearly, the intraorbital disorder has a negative correction to the model parameter $m_z$, provided $m_0>0$. On the contrary, the interorbital disorder gives rise to a positive correction to the parameter $m_z$. This explains why the intraorbital disorder and the interorbital disorder induce phase transitions in the opposite directions. The features of phase diagrams of the system are determined by the competition of intra- and interorbital disorders.

\section{Conclusion}
\label{Conclusion}
In summary, we investigate the combined effect of intra- and interorbital disorders in WSMs. By numerically calculating the normalized localization length and the Hall conductivity, we show rich phase diagrams due to the interplay of intra- and interorbital disorders. We found that the WSM with well-separated Weyl nodes is stable to weak intra- and interorbital disorder. The weak intraorbital disorder can induce a WSM-3D QAHI transition or a NI-WSM transition, while the interorbital disorder can cause phase transitions in inverse directions, i.e., a 3D QAHI-WSM transition and a WSM-NI transition. In the framework of the Born approximation, these phase transitions in the weak disorder regime can be well explained by the renormalization effect of the topological mass. We found that these two types of disorder give rise to corrections with opposite signs to the bare topological mass. We observe a 3D DAHM phase at moderate disorder strength. The DAHM appears in a wide range of the phase diagram when the intraorbital disorder dominates over the interorbital disorder, while the DAHM is found to exist in a narrow region or be missing in the phase diagram when the interorbital disorder dominates. Phase diagrams in $(W_{xy},W_z)$ space display distinguishing features compared with the phase diagram produced by the intraorbital disorder alone.

Recently, WSM phases induced by externally applied pressure and strain are reported \cite{Chang16ScienceAdv,LiuCC16PRL}. Meanwhile, it is found that externally applied pressure and strain can also drive WSM-NI phase transitions \cite{Chang16ScienceAdv,LiuCC16PRL}. Our study could be applied in these works since the lattice distortion induced by pressure or strain may be described by the intra- and interorbital disorders. Although the intra- and interorbital disorders are described by onsite potentials in our study, we believe they catch the essential physics in the localization and phase transitions. A more realistic way to model the lattice distortion is to introduce the random hopping type disorders and study the competition of the in-plane and out-of-plane random hopping disorders, and it will be explored in our future work.

In addition, it was shown that the on-site Hubbard interaction can annihilate a WSM and give rise to a 3D QAHI \cite{Laubach17PRB} in the present model we use, which means we can also manipulate different topological phases by tuning electron-election interaction. Therefore, we can expect that the interplay of disorder with electron-electron interaction \cite{Syzronov18ARCM} in WSM systems will produce more intriguing physical phenomena.

\section*{Acknowledgments}
R.C. and D.-H.X. were supported by the NSFC (Grant No. 11704106). D.-H.X. also acknowledges the support of Chutian Scholars Program in Hubei Province. B.Z. was supported by the NSFC (Grant No. 11274102), the Program for New Century Excellent Talents in University of Ministry of Education of China (Grant No. NCET-11-0960), and Specialized Research Fund for the Doctoral Program of Higher Education of China (Grant No. 20134208110001). J.-H.S. was supported in part by NSFC (under Grant No. 11604166) and the K.C. Wong Magna Fund in Ningbo University. C.-Z. C. thanks Juntao Song for illuminating discussions.


\begin{thebibliography}{99}
\bibitem{Hasan10RMP} M. Z. Hasan and C. L. Kane, Rev. Mod. Phys. \textbf{82}, 3045 (2010).

\bibitem{Qi11RMP} X.-L. Qi and S.-C. Zhang, Rev. Mod. Phys. \textbf{83}, 1057 (2011).

\bibitem{Wan11PRB}X. Wan, A. M. Turner, A. Vishwanath, and S. Y. Savrasov, Phys. Rev. B \textbf{83}, 205101 (2011).

\bibitem{Nielsen81PhysicsB1}H. B. Nielsen and M. Ninomiya, Nucl. Phys. B \textbf{185}, 20 (1981).

\bibitem{Nielsen81PhysicsB2}H. B. Nielsen and M. Ninomiya, Nucl. Phys. B \textbf{193}, 173 (1981).

\bibitem{XiongJ15Science} J. Xiong, S. K. Kushwaha, T. Liang, J. W. Krizan, M. Hirschberger, W. Wang, R. J. Cava, and N. P. Ong, Science \textbf{350}, 413 (2015).
\bibitem{HuangX15PRX} X. Huang, L. Zhao, Y. Long, P. Wang, D. Chen, Z. Yang, H. Liang, M. Xue, H. Weng, Z. Fang, X. Dai, and G. Chen, Phys. Rev. X \textbf{5}, 031023 (2015).

\bibitem{ZhangCL16NatComm} C.-L. Zhang, S.-Y. Xu, I. Belopolski, Z. Yuan, Z. Lin, B. Tong, G. Bian, N. Alidoust, C.-C. Lee, S.-M. Huang, T.-R. Chang, G. Chang, C.-H. Hsu, H.-T. Jeng, M. Neupane, D. S. Sanchez, H. Zheng, J. Wang, H. Lin, C. Zhang, H.-Z. Lu, S.-Q. Shen, T. Neupert, M.-Z. Hasan, and S. Jia, Nat. Commun. \textbf{7}, 10735 (2016).

\bibitem{Potter14NatComm}A. C. Potter, I. Kimchi, and A. Vishwanath, Nat. Commun. \textbf{5}, 5161 (2014).
\bibitem{Moll16Nature} P. J. W. Moll, N. L. Nair, T. Helm, A. C. Potter, I. Kimchi, A. Vishwanath, and J. G. Analytis, Nature \textbf{535}, 266 (2016).
\bibitem{DaiX16NatPhys} X. Dai, Nat. Phys. \textbf{12}, 727 (2016).
\bibitem{ZhangCL17NatPhys} C.-L. Zhang, S.-Y. Xu, C. M. Wang, Z. Lin, Z. Z. Du, C. Guo, C.-C. Lee, H. Lu, Y. Feng, S.-M. Huang, G. Chang, C.-H. Hsu, H. Liu, H. Lin, L. Li, C. Zhang, J. Zhang, X.-C. Xie, T. Neupert, M. Z. Hasan, H.-Z. Lu, J. Wang, and S. Jia, Nat. Phys. \textbf{13}, 979 (2017).
\bibitem{Roy16arXiv}B. Roy, R.-J. Slager, and V. Juricic, arXiv:1610.08973 (2016).

\bibitem{Hosur12PRL}P. Hosur, S. A. Parameswaran, and A. Vishwanath, Phys.
Rev. Lett. \textbf{108}, 046602 (2012).
\bibitem{Ominato14PRB}Y. Ominato and M. Koshino, Phys. Rev. B \textbf{89}, 054202
(2014).
\bibitem{Biswas14PRB}R. R. Biswas and S. Ryu, Phys. Rev. B \textbf{89}, 014205
(2014).
\bibitem{Sbierski14PRL}B. Sbierski, G. Pohl, E. J. Bergholtz, and P. W. Brouwer,
Phys. Rev. Lett. \textbf{113}, 026602 (2014).
\bibitem{Sbierski15PRB}B. Sbierski, E. J. Bergholtz, and P. W. Brouwer, Phys. Rev.
B \textbf{92}, 115145 (2015).

\bibitem{Baireuther14PRB}P. Baireuther, J. M. Edge, I. C. Fulga, C. W. J. Beenakker,
and J. Tworzyd\l o, Phys. Rev. B \textbf{89}, 035410 (2014).

\bibitem{Nandkishore14PRB}R. Nandkishore, D. A. Huse, and S. L. Sondhi, Phys. Rev. B
\textbf{89}, 245110 (2014).
\bibitem{Kobayashi2014}K. Kobayashi, T. Ohtsuki, K.-I. Imura, and I. F. Herbut, Phys. Rev. Lett. \textbf{112}, 016402 (2014).
\bibitem{Pixley15PRL}J. H. Pixley, P. Goswami, and S. Das Sarma, Phys. Rev. Lett.
\textbf{115}, 076601 (2015).
\bibitem{Pixley16PRB}J. H. Pixley, P. Goswami, and S. Das Sarma, Phys. Rev. B
\textbf{93}, 085103 (2016).
\bibitem{Pixley16PRX}J. H. Pixley, D. A. Huse, and S. Das Sarma, Phys. Rev. X \textbf{6}, 021042 (2016).
\bibitem{Syzronov15PRL}S. V. Syzranov, L. Radzihovsky, and V. Gurarie, Phys. Rev. Lett. \textbf{114}, 166601 (2015).
\bibitem{Syzronov18ARCM}S. V. Syzranov, and L. Radzihovsky, Annu. Rev. Condens. Matter Phys. \textbf{9}, 35 (2018).
\bibitem{Altland15PRL}A. Altland and D. Bagrets, Phys. Rev. Lett. \textbf{114}, 257201
(2015).

\bibitem{Shapourian16PRB}H. Shapourian and T. L. Hughes, Phys. Rev. B \textbf{93}, 075108 (2016).

\bibitem{Liu16PRL}S. Liu, T. Ohtsuki, and R. Shindou, Phys. Rev. Lett. \textbf{116}, 066401 (2016).

\bibitem{Chen15PRL}C.-Z. Chen, J. Song, H. Jiang, Q.-f. Sun, Z. Wang, and X. C. Xie, Phys. Rev. Lett. \textbf{115}, 246603 (2015).
\bibitem{Bera16PRB}S. Bera, J. D. Sau, and B. Roy, Phys. Rev. B \textbf{93}, 201302 (2016).

\bibitem{Park17PRB}M. J. Park, B. Basa, and M. J. Gilbert, Phys. Rev. B \textbf{95}, 094201 (2017).

\bibitem{Wu17PRB}Y. Wu, H. Liu, H. Jiang, and X. C. Xie, Phys. Rev. B \textbf{96}, 024201 (2017).

\bibitem{Song12PRB}J. Song, H. Liu, H. Jiang, Q.-f. Sun, and X. C. Xie, Phys. Rev. B \textbf{85}, 195125 (2012).
\bibitem{Hu16PRB} L. H. Hu, D. H. Xu, F. C. Zhang, and Y. Zhou, Phys. Rev. B \textbf{94}, 085306 (2016).
\bibitem{QiaoZ16PRL}Z. Qiao, Y. Han, L. Zhang, K. Wang, X. Deng, H. Jiang, S. A. Yang, J. Wang, and Q. Niu, Phys. Rev.
Lett. \textbf{117}, 056802 (2016).
\bibitem{Hung16PRB} H.-H. Hung, A. Barr, E. Prodan, and G. A. Fiete, Phys. Rev. B \textbf{94}, 235132 (2016).

\bibitem {LiJ09PRL}J. Li, R. L. Chu, J. K. Jain, and S. Q. Shen, Phys. Rev.
Lett. \textbf{102}, 136806 (2009).

\bibitem {Jiang09PRB}H. Jiang, L. Wang, Q.-f. Sun, and X. C. Xie, Phys. Rev. B
\textbf{80}, 165316 (2009)
\bibitem {Groth09PRL}C. W. Groth, M. Wimmer, A. R. Akhmerov, J. Tworzyd\l o, and C. W. J. Beenakker, Phys. Rev. Lett. \textbf{103}, 196805 (2009).
\bibitem{Jiang16cpb}B. Wu, J. Song, J. Zhou, and H. Jiang, Chin. Phys. B \textbf{11}, 117311 (2016).
\bibitem{Yang11PRB}K.-Y. Yang, Y.-M. Lu, and Y. Ran, Phys. Rev. B \textbf{84}, 075129 (2011).

\bibitem{MacKinnon81PRL}A. MacKinnon and B. Kramer, Phys. Rev. Lett. \textbf{47}, 1546 (1981).

\bibitem{MacKinnon83PhysB}A. MacKinnon and B. Kramer, Z. Phys. B: Condens. Matt. \textbf{53}, 1 (1983).

\bibitem{Kramer93RPP}B. Kramer and A. MacKinnon, Rep. Prog. Phys. \textbf{56}, 1469 (1993).

\bibitem{Nagaosa03PRL}M. Onoda and N. Nagaosa Phys. Rev. Lett. \textbf{90}, 206601 (2003).
\bibitem{Slevin04PRB} K. Slevin, Y. Asada, and L. I. Deych, Phys. Rev. B \textbf{70}, 054201 (2004).
\bibitem{Onoda07PRL}M. Onoda, Y. Avishai, and N. Nagaosa, Phys. Rev. Lett. \textbf{98}, 076802 (2007).

\bibitem{Yamakage13PRB}A. Yamakage, K. Nomura, K.-I. Imura, and Y. Kuramoto, Phys. Rev. B \textbf{87}, 205141 (2013).

\bibitem{Kobayashi13PRL}K. Kobayashi, T. Ohtsuki, and K.-I. Imura, Phys. Rev. Lett. \textbf{110}, 236803 (2013).

\bibitem{Chen15PRB}C.-Z. Chen, H. Liu, H. Jiang, Q.-f. Sun, Z. Wang, and X. C. Xie, Phys. Rev. B \textbf{91}, 214202 (2015).

\bibitem{Prodan11JPA}E. Prodan, J. Phys. A \textbf{44}, 113001 (2011).

\bibitem{Prodan13AMRE}E. Prodan, Appl. Math. Res. Express \textbf{2013}, 176 (2013).

\bibitem{Song14PRB1}J. Song and E. Prodan, Phys. Rev. B \textbf{89}, 224203 (2014)

\bibitem{Song14PRB2}J. Song, C. Fine, and E. Prodan, Phys. Rev. B \textbf{90}, 184201 (2014).

\bibitem{Burkov11PRL}A. A. Burkov and L. Balents, Phys. Rev. Lett. \textbf{107}, 127205 (2011).
\bibitem{Burkov11PRB}A. A. Burkov, M. D. Hook, and L. Balents, Phys.
Rev. B \textbf{84}, 235126 (2011).

\bibitem{Chang16ScienceAdv}G. Chang, S.-Y. Xu, D. S. Sanchez, S.-M. Huang, C.-C. Lee, T.-R. Chang, G. Bian, H. Zheng, I. Belopolski, N. Alidoust, H.-T. Jeng, A. Bansil, H. Lin, and M. Z. Hasan, Sci. Adv. \textbf{2}, e1600295 (2016).

\bibitem{LiuCC16PRL}C.-C. Liu, J.-J. Zhou, Y. Yao, and F. Zhang, Phys. Rev. Lett. \textbf{116}, 066801 (2016).
\bibitem{Laubach17PRB} M. Laubach, C. Platt, R. Thomale, T. Neupert, and S. Rachel, Phys. Rev. B \textbf{94}, 241102(R) (2016).
\end{thebibliography}
\end{document}